\documentstyle[preprint,aps,epsf]{revtex}
\begin{document}
\input psfig.tex
\draft

\title{Are Protein Folds Atypical?}
\author{Hao Li, Chao Tang\cite{tang}, and Ned S. Wingreen}
\address{NEC Research Institute, 4 Independence Way, Princeton, New
Jersey 08540}

\date{August 29, 1997}
\maketitle

\begin{abstract}
Protein structures are a very special class among all possible
structures.  It was suggested that a ``designability principle'' plays a
crucial role in nature's selection of protein sequences and structures.
Here we provide a theoretical base for such a selection principle, using
a novel formulation of the protein folding problem based on hydrophobic
interactions.  A structure is
reduced to a string of 0's and 1's which represent the surface and core
sites, respectively, as the backbone is traced.  Each structure is
therefore associated with one point in a high dimensional space.
Sequences are represented by strings of their hydrophobicities and thus
can be mapped into the same space.  A sequence which lies closer to a
particular structure in this space than to any other structures will
have that structure as its ground state.  Atypical
structures, namely those far away from other structures in the high
dimensional space, have more sequences which fold into them, and are
thermodynamically more stable.  We argue that the most common folds of
proteins are the most atypical in the space of possible structures.
\end{abstract}

\narrowtext


Protein structures seem to be a very special class among all possible
folded configurations of a polypeptide chain.  There are preferred
secondary structures and motifs \cite{lev} as well as striking
regularities in the geometries of protein structures \cite{ric1,ric2}. 
Two proteins are said to have a common fold if they have the same major
secondary structures in the same arrangement with the same topological
connections \cite{mur}.  Common folds occur even for proteins with
different biological functions.  Indeed there are ``superfolds''
\cite{ore} and ``fold space attractors'' \cite{hol} which account for
the structures of many proteins.  It has been estimated that the total
number of natural protein folds is only about 1000 \cite{cho}, an
extremely small number compared with both the number of proteins and
the number of all possible structures.
One may ask: is there anything special about natural protein structures
-- are they merely an arbitrary outcome of evolution or is there some
fundamental reason behind their selection?  

Many features of folded proteins can be understood from an energetic
point of view \cite{cf}.  Close packing of secondary structures favors
some geometrical patterns \cite{clr}.  It was argued that
certain empirical rules for connections between secondary structures can
be explained by lower bending energies for the connecting loops \cite{fp}.
It was further conjectured that a structure with lower energy would
stabilize more sequences and would be more likely a protein fold
\cite{fgb}.  However, the number of low energy structures is 
huge, and overall low energy does not necessarily imply
thermodynamic stability since other low energy structures still compete
with the ground state.  Clearly, purely energetic considerations are
not the whole story.

A recent study on a simple model of protein folding suggested that a
rather different mechanism -- the so-called ``designability principle''
-- should play a crucial role in nature's selection of protein sequences
and structures \cite{lhtw}.  The designability of a structure is defined
as the number of sequences that possess the structure as their
nondegenerate ground state.  It was demonstrated in the model that
structures with the same low energy (when averaged over sequences) differ
drastically in terms
of their designability; highly designable structures emerge with a number
of associated sequences much larger than the average.  These highly
designable structures are relatively stable against mutation and are more
thermodynamically stable than other structures.  In addition, they possess 
``proteinlike'' secondary structures and motifs.  A number of questions
arise: Among the large number of low energy structures, why are some
structures highly designable?  Why does designability also guarantee
thermodynamic stability?  Why do highly designable structures have
geometrical regularities?  Here we address these questions using a novel
formulation of protein folding problem based on hydrophobic interactions.

Among the various forces involved in the folding of a
polypeptide chain -- van der Waals force, electrostatic force, hydrogen
bonding, hydrophobic force -- there is strong and increasing
evidence that the hydrophobic force is the dominant one in 
determining the overall folded structure \cite{kau,dil}.  The hydrophobic
force originates from the contact of nonpolar groups with water, which
disrupts the hydrogen bonding exchange pattern between water molecules.
Thus nonpolar groups in water tend to coalesce to minimize their contact
with water.  For a nonpolar amino acid, the free energy reduction
from the hydrophobic interaction is proportional to the total area of the
side chain protected from water \cite{cho2,her,rich,eis}.  For an amino
acid with
a polar side chain, there is a smaller reduction because of the possibility
of hydrogen bonding between the polar side chain and water molecules
\cite{cho2}.  To model the hydrophobic force in protein folding, one can
assign parameters $h_\sigma$ to characterize the hydrophobicities of each
of the 20 amino acids \cite{eis}.  Each sequence of amino acids then has an
associated vector 
${\bf h}=(h_{\sigma_1},h_{\sigma_2},\ldots,h_{\sigma_i},\ldots,h_{\sigma_N})$,
where $\sigma_i$ specifies the amino acid at position $i$ of the sequence. 
We take the energy of a sequence folded into a particular structure to be
the sum of the contributions from each amino acid upon burial away from water: 
\begin{equation}
H=-\sum_i s_i h_{\sigma_i},
\label{ham1}
\end{equation}
where $s_i$ is a structure-dependent number characterizing the degree of
burial of the $i$-th amino acid in the peptide chain \cite{ltw}.
Here we ignore all other forces which, undoubtedly, help determine the details
of a protein's structure.  The advantage of considering only the hydrophobic
force is that it drastically simplifies the analysis and thereby elucidates
some essential features of the folding problem. 

\begin{figure}
\vskip 0.5true cm
\centerline{\epsfxsize=4in
\epsffile{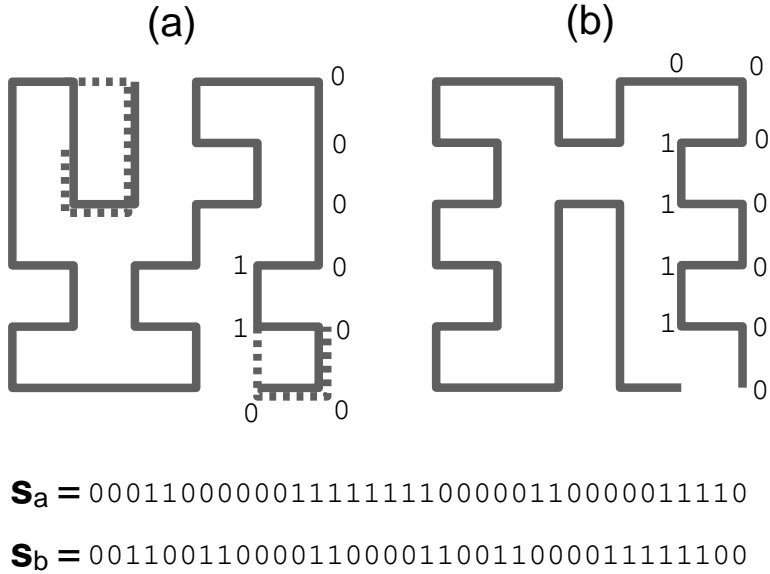}}
\vskip 0.5true cm
\caption{Structures are represented by strings ${\bf s}$ of 0's and 1's,
according to whether a site is on the surface or in the core,
respectively.  Shown are two examples of compact $6\times6$ lattice
structures.
(a) A typical structure. Dotted lines indicate local changes that can
be performed to transform it to other compact structures.  Note that the
change at the lower right corner does not change the string pattern, so
this structure is a degenerate one. (b) The most designable structure.}
\label{struct}
\end{figure}

To simplify the application of Eq.~(\ref{ham1}), let us consider only
globular compact structures and let $s_i$ take only two values: 0 and 1,
depending on whether the amino acid is on the surface or in the core 
of the structure, respectively.  Therefore, each compact structure can
be represented by a string $\{s_i\}$ of 0's and 1's: $s_i=0$ if the $i$-th
amino acid is on the surface and $s_i=1$ if it is in the core (see
Fig.~\ref{struct} for an example on a lattice).  Assuming every compact
structure of a given size has the same numbers of surface and core sites
and noting that the term $\sum_i h_{\sigma_i}^2$ is a constant for a
fixed sequence of amino acids and does not play any role in determining
the relative energies of structures folded by the sequence, 
Eq.~(\ref{ham1}) is equivalent to:
\begin{equation}
H = \sum_{i=1}^N (h_{\sigma_i} - s_i)^2.
\label{ham2}
\end{equation}

Having formulated the protein folding problem in terms of
Eq.~(\ref{ham2}), we now proceed to make a few observations.
The problem involves two spaces: the sequence space
and the structure space.  We represent a sequence by the vector of its
hydrophobicities ${\bf h}=(h_{\sigma_1},h_{\sigma_2},\ldots,h_{\sigma_N})$,
and the sequence space $\{{\bf h}\}$ consists of $20^N$ sequences since
there can be any of 20 amino acids at each site.  A structure is also
represented by a vector ${\bf s}=(s_1,s_2,\ldots,s_N)$, where $s_i=0$ or
1, and the structure space $\{{\bf s}\}$ consists of the $2^N$ possible
strings of 0's and 1's.  However, only a small subset of the strings of
0's and 1's represent realizable structures.  If two or more structures
map into the same string, we say that these structures are degenerate (see
Fig.~\ref{struct}(a)).  It is evident that a degenerate structure cannot
be the unique ground state for any sequence within this formulation.  The
fraction of all structures which are nondegenerate depends on the ratio
of surface sites to core sites.  This fraction approaches zero in the
limits of very large and very small surface-to-core ratios.  It is
worthwhile noting that for natural proteins the surface-to-core ratio is
of the order one. 

\begin{figure}
\centerline{\epsfxsize=3.3in
\epsffile{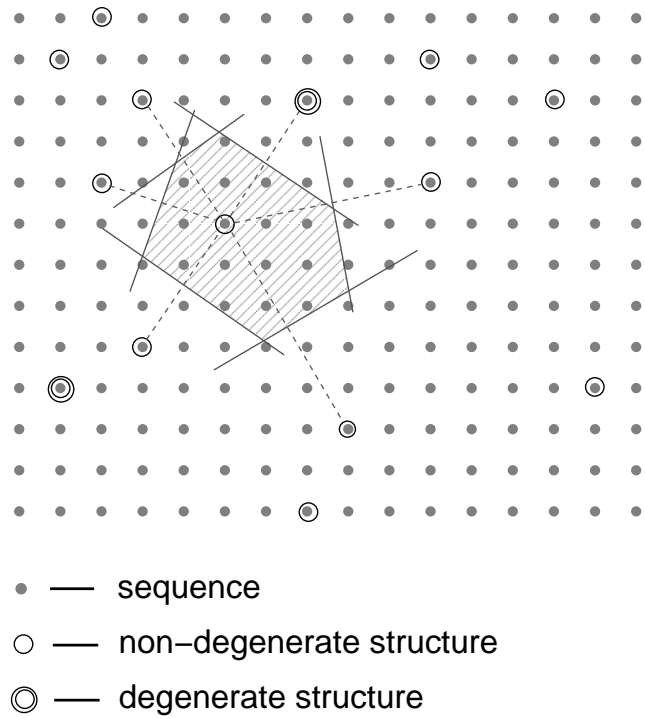}}
\vskip 0.5true cm
\caption{Schematic plot of the sequence and the structure spaces and the
Voronoi construction.  The Voronoi polytope is the shaded region.}
\label{voron}
\end{figure}

Now imagine embedding both the sequence space $\{{\bf h}\}$ and the
structure space $\{{\bf s}\}$ in an $N$-dimensional euclidean space
(Fig.~\ref{voron}).  This is simplest to picture if one normalizes the
$h_\sigma$ so that $0\le h_\sigma \le 1$.  Since the energy for a 
sequence ${\bf h}$ folded
into a structure ${\bf s}$ is the square of the distance between ${\bf
h}$ and ${\bf s}$ (Eq.~(\ref{ham2})), it is evident that ${\bf h}$ will
have ${\bf s}$ as its unique ground state if and only if ${\bf h}$ is
closer to ${\bf s}$ than to any other structure.  Therefore, the set of
all sequences $\{{\bf h}({\bf s})\}$
which uniquely design a structure ${\bf s}$ can be found by the following
geometrical construction: Draw bisector planes between ${\bf s}$ and all
of its neighboring structures in the $N$-dimensional space (see 
Fig.~\ref{voron}).  The volume enclosed by these planes is called
the Voronoi polytope around ${\bf s}$.  $\{{\bf h}({\bf s})\}$ then
consists of all sequences within the Voronoi polytope.  Hence, the
designabilities of structures are directly related to the distribution
of the structures in the structure space $\{{\bf s}\}$.  A structure
closely surrounded by many neighbors will have a small Voronoi polytope
and hence a low designability; while a structure far away from others will
have a large Voronoi polytope and hence a high designability.  Furthermore,
the thermodynamic stability of a folded structure is directly related to
the size of its Voronoi polytope.  For a sequence ${\bf h}$, the energy
gap between the ground state and an excited state is the difference of
the squared distances between ${\bf h}$ and the two states
(Eq.~(\ref{ham2})).  A larger Voronoi polytope implies, on average, a
larger gap as excited states can only lie outside of the Voronoi polytope
of the ground state.  Thus, this geometrical representation of the problem
naturally explains the positive correlation between the thermodynamic
stability and the designability, an observation made in Ref.~\cite{lhtw}.

\begin{figure}
\centerline{\epsfxsize=4in
\epsffile{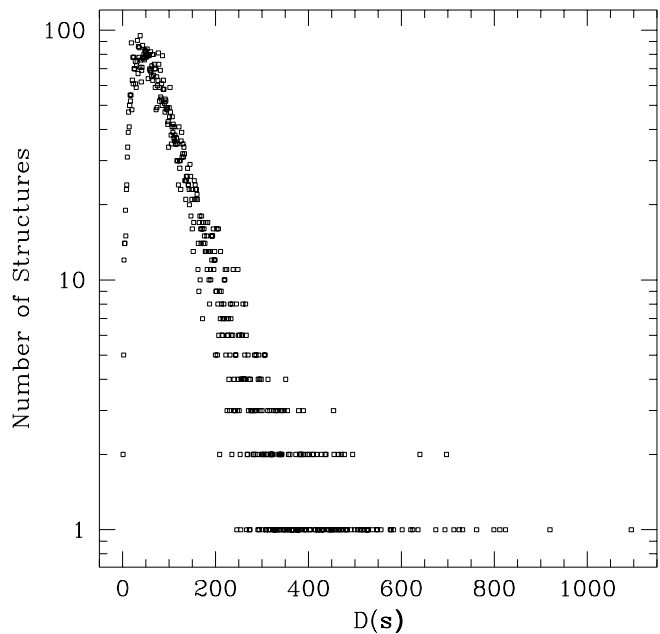}}
\vskip -0.2true cm
\caption{Histogram of the designability obtained by random sampling using
19,492,200 sequences.}
\label{histo}
\end{figure}

To further illustrate and elaborate on the above ideas, let us proceed with
a simple example: a two-dimensional lattice HP model \cite{hp}.  Instead of
20, we use only two amino acids: H (hydrophobic) and P (polar).  The vector 
representing a sequence is now ${\bf h}=(h_1,h_2,\ldots,h_i,\ldots,h_N)$,
where $h_i=1$ if the $i$-th amino acid is an H and $h_i=0$ if it is a P. 
The sequence space now coincides with the structure space, both
consisting of all the possible strings of 0's and 1's of length $N$. 
To obtain a set of allowed structure strings, we focus on the compact
$6\times6$ two-dimensional lattice structures (Fig.~\ref{struct}),
which can be easily enumerated.  We divide the
36 sites into 20 surface sites and 16 core sites; the surface/core ratio
is 1.25.  There are 57,337 compact structures not related by symmetries.
These structures map into 30,408 distinct strings, among which 
18,213 ($\sim 30\%$ of all structures) represent nondegenerate structures. 
To obtain a histogram of the designability for all structures, we randomly
sampled the sequence space.  We randomly chose $\sim$20 million sequences
(enough to suppress statistical fluctuations), and for each of these 
calculated its energy on all the structure strings using Eq.~(\ref{ham1}).
We found that $\sim 8.8\%$ of all
the sequences chosen have unique ground states, i.e., they fall inside
of the Voronoi polytope of some nondegenerate structure.  The designability
histogram obtained in the sampling is plotted in Fig.~\ref{histo}.
Recall that the designability of a structure $D({\bf s})$ is the number
of sequences found to have that structure as their unique ground state.
The designabilities obtained from this model are very well correlated
with those found in the previous work \cite{lhtw} which were obtained
from a different HP model, and with those calculated using the
Miyazawa-Jernigan interaction matrix \cite{miy} for all 20 amino acids.
The sets of highly designable structures are essentially the same for all
three models.  

Certain features of Fig.~\ref{histo} can be understood from the following
simple consideration.  If the structure strings were {\it randomly}
distributed in structure space, then one would expect the distribution
of the volumes $V$ of the Voronoi polytopes constructed around every
structure to be $\propto \exp(-cV)$, for $cV \gg 1$, where $c$ is
proportional to the overall density of structure strings in structure
space \cite{gil}.  Moreover, the probability of finding very small
Voronoi volumes $V$ would also vanish because small $V$'s require a
close clustering of randomly distributed points \cite{vor}.
The distribution shown in
Fig.~\ref{histo} indeed has a rising part for small designability $D$
and a decaying part which is roughly exponential.  However, one should not
conclude from this that the structures are randomly distributed in 
structure space.  In fact, the structures are highly correlated and
clustered.  For example, in Fig.~\ref{histo} there are
structures (with $D({\bf s})>500$) which are highly designable and which
do not follow the exponential decay.  

\begin{figure}
\centerline{\epsfxsize=4in
\epsffile{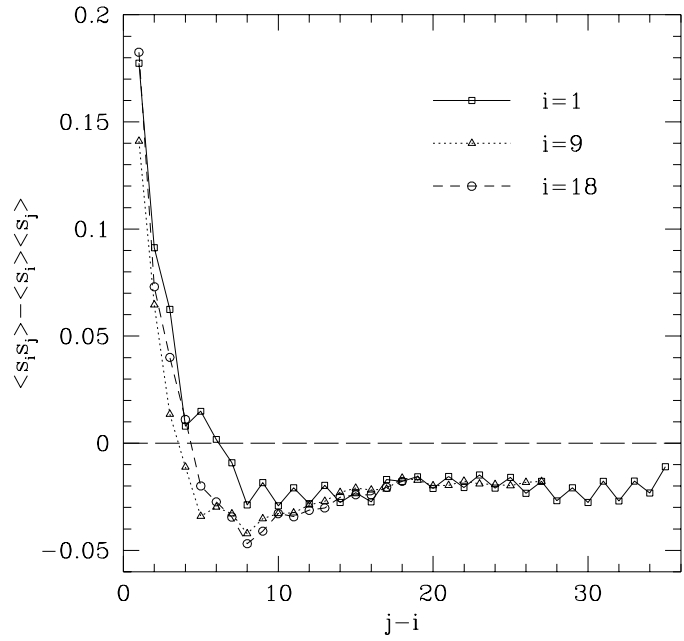}}
\caption{Correlation functions for the structures.}
\label{corr}
\end{figure}

In order to more fully understand the designabilities of structures, one
has to know the distribution of structures in structure space.  Each
realizable string ${\bf s}$ represents one (sometimes more than one)
globular compact structure; its components $s_i$ are determined by the
geometry of the represented structure.  One trivial correlation between
the realizable ${\bf s}$'s comes from the constraint that $\sum_i
s_i=n_c$, where $n_c$ is the number of core sites and is a constant
depending only on $N$.  Therefore the allowed structures reside in a
subspace consisting of those strings whose Hamming distance \cite{hamm}
to the origin equals $n_c$. 
Other more subtle correlations are related to the properties of
compact self-avoiding walks of length $N$.  To characterize the ensemble
properties of the allowed structures, we measure the correlation function
$c(i,j)=<s_is_j>-<s_i><s_j>$, where the average is taken over all 57,337
compact structures.  In Fig.~\ref{corr}, $c(i,j)$ is plotted vs. $j-i$ for
three different starting positions: $i=1$, $i=9$, and $i=18$.  A common
feature of these correlation functions is that $s_i$ has positive
correlation with nearby sites and that the correlation turns negative as
$j-i$ increases.  This feature simply reflects a property of self-avoiding
random walks in a confined geometry: if a site is in the core then most
likely the next few sites are also in the core, and similarly if a site is
on the surface the next few sites are most likely on the surface.  From
Fig.~\ref{corr} we see that the transition between core and surface takes,
on average, about 4 to 6 steps.  The residual negative correlation at
large distances is a result of the constraint that there is a fixed number
of core (and surface) sites.

\begin{figure}
\centerline{\epsfxsize=4in
\epsffile{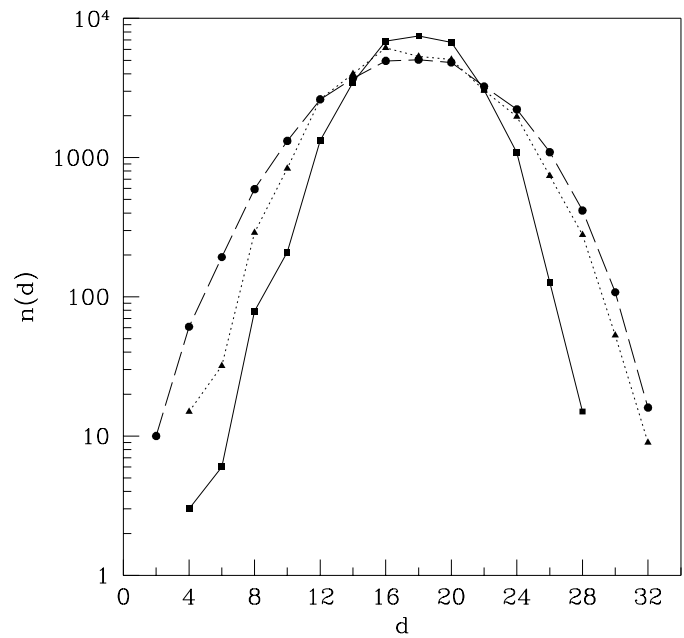}}
\caption{Number of structures vs. the Hamming distance for three typical
structures with: low (circles), intermediate (triangles), and high
(squares) designability.}
\label{nd}
\end{figure}

Another way to measure the correlation of the allowed structures in
structure space is to measure the number of distinct structure strings, 
$n_{\bf s}(d)$, at a Hamming distance $d$ from a given string ${\bf s}$. 
In Fig.~\ref{nd}, $n_{\bf s}(d)$ is plotted for three different structures
with low, intermediate, and high designabilities, respectively.  
A highly designable structure typically has fewer neighbors than a less
designable structure, not only at the smallest $d$'s but out to $d$'s of
order 10 - 12.  This indicates that there exists a local ``density'' around
each allowed structure in the structure space. 

From the above discussion, we see that structure strings are not
randomly distributed in the structure space.  Their distribution is
highly correlated.  There are high density regions as well as regions
with very low density.  High density regions consist of ``typical''
structures or typical random walks (see Fig.~\ref{struct}(a) for an
example) whose correlation functions are similar to the correlation
function of the ensemble (Fig.~\ref{corr}).  These structures usually
have small Voronoi polytopes because they are closely surrounded by many
neighbors.  ``Atypical'' or ``rare'' structures reside in regions with
very low local density, and their correlation functions are generally
very different from Fig.~\ref{corr}.
Atypical structures usually have high designabilities because they are
relatively far away from other structures in the structure space.  One
way for a structure to have an atypical correlation function is to
have many surface to core transitions along its backbone.  This leads to
geometrically regular pleated patterns such as the one shown in
Fig.~\ref{struct}(b), which is in fact the most designable structure.
Thus, the emergence of the ``proteinlike'' sub-structures and motifs in
the highly designable structures \cite{lhtw} occurs not because they
are common, but precisely because they are rare.  A related property of
rare structures, which follows from the large Voronoi polytope, is that
it is hard to make any local change to them.  For the ``typical''
structure in Fig.~\ref{struct}(a), one can make local changes as
indicated by the dotted lines to transform the structure to other
compact structures.  It is impossible to perform any such local changes
for the most designable -- and therefore most ``atypical'' -- structure
in Fig.~\ref{struct}(b). 

In conclusion, we have formulated the protein folding problem in a novel
way using only hydrophobic interactions.  A very simple and
transparent picture emerges from the formulation in which the
designability of a structure, i.e., the number of sequences for which the
structure is a unique ground state, is directly related to the volume of the
Voronoi polytope around the structure.  Structures with atypical
patterns of surface and core sites have high designability and are
generally more stable thermodynamically.  These atypical patterns
produce geometrically regular structures with ``proteinlike'' motifs. 
According to this picture, the most common folds in natural
proteins are the most atypical in the space of possible structures.
Note that highly designable structures do not have lower energies than
other structures.  It is easy to see from Eq.~(\ref{ham1}) that the
average energy and the energy spectrum (over all the sequences) are the
same for all the compact globular structures.  Thus a structure is highly
designable in our model, not because it has a lower energy or an unusual
energy spectrum as conjectured by Finkelstein and colleagues \cite{fgb},
but because it is far away from neighboring competing structures in the
structure space.  In our discussion we considered only the hydrophobic
interaction and compact globular structures.  The formulation can be
generalized, at least conceptually, without much difficulty to include
non-compact structures and main-chain hydrogen bonding, both of which
are purely structural properties.  The overall picture is unchanged.
Our results can be tested on real protein structures.  For a globular
protein there is a natural division into core and surface sites and thus a
string representation.  The strings of natural proteins can be compared
with the ensemble of strings of random structures.  The knowledge of the
distribution of allowed structures in structure space can be used to aid
protein structure classification, protein design, and structure prediction.

We thank Robert Helling for helpful discussions.


\end{document}